\documentclass{aastex}
\usepackage{spr-astr-addons}
\usepackage{url}

\RequirePackage{color}

\begin{document}

\title{	Classification of BeppoSAX's Gamma-Ray Bursts}
\shorttitle{Classification of BeppoSAX's GRBs}
\shortauthors{I. Horv\'ath}

\author{Istv\'an Horv\'ath \altaffilmark{1}}


\altaffiltext{1}{Dept. of Physics,
              Bolyai Military University, Budapest,
              POB 15, H-1581, Hungary\\
              \email{horvath.istvan@zmne.hu}
}

\begin{abstract}

The BeppoSAX Catalog has been very recently published.
In this paper we analyze - using the Maximum
Likelihood (ML) method - the duration distribution
of the 1003 GRBs listed in the catalog with duration.
The ML method can identify the long and the intermediate
duration groups. The short population of the bursts
is identified only at a 96\% significance level.
MC simulation has been also applied and 
gives a similar significance level;  95\%. 
However, the existence of the short bursts is
not a scientific question after the Compton Gamma-Ray
Observatory's observation.
Our minor result is this 
well-known fact that in the BeppoSAX
data the short bursts are underrepresented,
mainly caused by the different triggering system.
Our major result is the identification
of the intermediate group in the BeppoSAX data.
Therefore, four different satellites (CGRO, Swift, RHESSI and
BeppoSAX) observed the intermediate type Gamma-Ray Burst.

\end{abstract}

\keywords{Gamma rays: bursts, theory, observations }


\section{Introduction}

\citep{kou93} identified two types of Gam\-ma-Ray Bursts (GRBs) based on durations, 
for which the value of $T_{90}$ (the time during which 90\% of the 
fluence is accumulated) is smaller or larger than 2 s, respectively.
\cite{muk98} and \citep{ho98} made a suggestion about
the third type of GRBs, which have intermediate (2-9 s)
duration.
Later several papers confirmed this suggestion
\citep{hak00,bala01,rm02,ho02,hak03,ho06,ch07}.
All of these works were based on BATSE observations.

Lately, analysis of Swift \citep{ho08} and RHESSI
\citep{rip07,rip09} data have 
 found a similar group structure
(for RHESSI the $\chi ^2$ method found only two groups \citep{rip07},
however, the Maximum Likelihood (ML) method
reveals a significant intermediate group \citep{rip09}).

Since The Gamma-Ray Burst catalog obtained with the 
Gamma Ray Burst Monitor (GRBM) aboard BeppoSAX \cite{fr08} has been published
recently, this catalog is worth to be analyzed,
whether the statistical methods can show us
similar or different groups of GRBs.

In Sect. 2 we
analyze the duration distribution of
the GRBs observed by BeppoSAX.
In Sect. 3 we discuss the instrumentational
details of our analysis.
In Sect. 4 we present the conclusions.

\section{Analysis of the duration distribution}

In the BeppoSAX catalog  \cite{fr08,fr09} there are
1082 GRBs, of which 1003 have duration information
 (see Figure 1. for the distribution).

\begin{figure}
\centering
\resizebox{\hsize}{!}
{\includegraphics[angle=0,width=7cm]{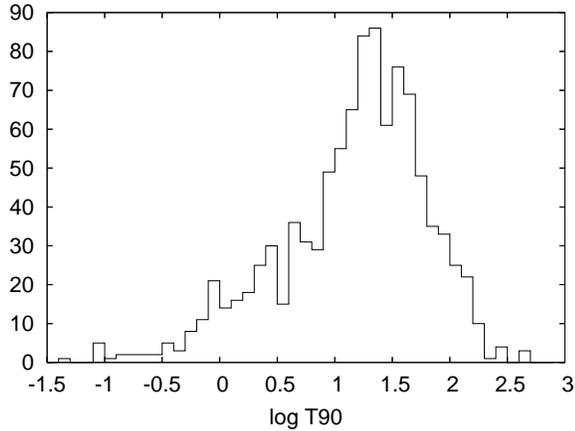}}
\caption{The duration distribution of the BeppoSAX 1003 bursts.  }
\end{figure}

\begin{figure}
\centering
\resizebox{\hsize}{!}
{\includegraphics[angle=0,width=7cm]{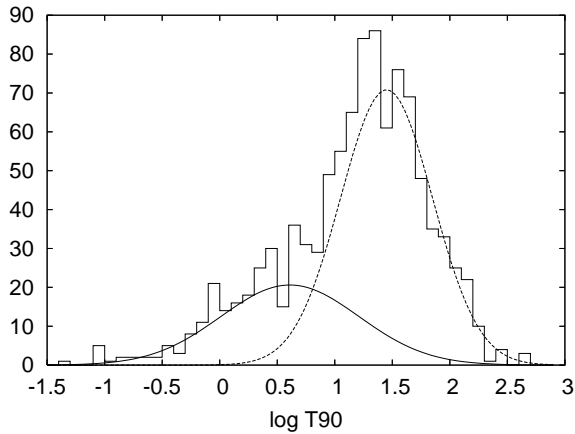}}
\caption{The duration distribution of the BeppoSAX bursts
with the Gaussian functions (long bursts with dashed line).  }
\end{figure}

\subsection{Maximum Likelihood calculations}

Similarly to \cite{ho02} we fit the $\log T_{90}$  distribution using Maximum Likelihood
(ML) method with a superposition of $k$ log-normal components, each
of them having 2 unknown parameters to be fitted with $N=1003$
measured points in our case. Our goal is to find the minimum value
of $k$ suitable to fit the observed distribution. Assuming a
weighted  superposition of $k$ log-normal distributions one has to
maximize the following likelihood function:

\begin{equation}
L_k = \sum_{i=1}^{N} \log  \left(\sum_{l=1}^k   w_lf_l(x_i,\log
T_l,\sigma_l ) \right)
\end{equation}

\noindent where $w_l$ is a weight, $f_l$ a log-normal function with
$\log T_l$ mean and $\sigma_l $ standard deviation having the form
of

\begin{equation}
$$f_l = \frac{1}{ \sigma_l  \sqrt{2 \pi  }}
\exp\left( - \frac{(x-\log T_l)^2}{2\sigma_l^2} \right)  $$
\label{fk}
\end{equation}

\noindent and due to a normalization condition

\begin{equation}\label{wight}
    \sum_{l=1}^k w_l= N \, .
\end{equation}

 \noindent We used a simple
C++ code to find the maximum of $L_k$. Assuming only one log-normal
component the fit gives $L_{1max}=5951.895$ but in the case of
$k$=2 one gets $L_{2max}=6011.355$.
 See Table 1. for the best parameters.

\begin{table}
\caption[]{The best parameters for the two log-normal fit of
the GRB (observed by BeppoSAX) duration distribution. } \label{2g}
$$
         \begin{array}{cccc}
            \hline
            \noalign{\smallskip}
     & Duration (log T_{90})  &  \sigma  (log T_{90}) &  w \\
            \noalign{\smallskip}
            \hline
            \noalign{\smallskip}
        shorter      &   0.62   &  0.62   &   306.2   \\
        long       &     1.45   &  0.40   &   696.8   \\
            \noalign{\smallskip}
            \hline
         \end{array}
     $$
\end{table}

In \cite{ho08} we summarize the ML method and
refer to \cite{KS76} the confidence region of the estimated parameters, which is given by the
following formula, where $L_{max}$ is the maximum value of the
likelihood function and $L_0$ is the
likelihood function  at the true value of the
parameters:

\begin{equation}
2 ( L_{max} -  L_0) \approx \chi^2_k \label{eq:chi}
\end{equation}

Based on this equation we can infer whether the addition of a
further log-normal component is necessary to significantly improve
the fit. We make the null hypothesis that we have already reached 
the true value of $k$. Adding a new component, i.e. moving
from $k$ to $k+1$, the ML solution of $L_{kmax}$ has changed to
$L_{(k+1)max}$, but $L_0$ remained the same. In the meantime we
increased the number of parameters with 3 ($w_{k+1}$, $logT_{k+1}$
and $\sigma_{(k+1)})$. Applying Eq. (\ref{eq:chi}) on both
$L_{kmax}$ and $L_{(k+1)max}$ we get after subtraction

\begin{equation}\label{kk1}
2 ( L_{(k+1)max} -  L_{kmax}) \approx \chi^2_3 \, .
\end{equation}

\noindent For $k=1$ $L_{2max}$ is greater than $L_{1max}$
by more than 59, which gives for $\chi^2_3$ an
extremely low probability. It means the
two log-normal fit is a really better approximation for
the duration distribution of GRBs than one log-normal.
The longer duration group (with a centroid about 28 seconds)
has 69\% of the burst population and the shorter one
(with a centroid about 4.2 seconds) has 31\%.

Thirdly, a
three-log-normal fit was made combining three $f_k$ functions with
eight independent parameters 
(three means, three standard deviations and two
weights).  The highest value of the
logarithm of the likelihood ($L_{3max}$) is 6015.585.  For two
log-normal functions the maximum  was $L_{2max}=6011.355$.  The maximum thus
improved by 4.23. Twice of this is 8.46 which gives us the
probability of 3.7\% for the difference between $L_{2max}$
and $L_{3max}$ is being only by chance. 

\subsection{Monte Carlo simulations}

One can check the 0.037 probability, using a Monte-Carlo (MC)
simulation.  Take the
two-log-normal distribution with the best fitted parameters of the
observed data, and generate 1003 numbers for $T_{90}$ whose
distribution follows the best fit of the two-log-normal distribution.  
Find the best likelihood with five free parameters, two means, two
dispersions and two weights (only one is independent). 
Make a fit with a three-log-normal distribution
(eight free parameters, three means, three dispersions and two  
independent weights).
Take the difference between the two logarithms of the
maximum likelihoods and that gives one number in our MC simulation.

This procedure have been carried out for 1000 simulations. 
There were  49 cases when the log-likelihood difference
was more than the one obtained for the BeppoSax data (4.23). 
Therefore, the MC simulations do not confirm exactly the numerical result 
obtaining by Eq. (\ref{kk1}), however, it gives 
a similar (4.9\%) probability for the statement that a 
third group is only a statistical fluctuation
(for more details see \cite{da03,da04}).

The conclusion is the same;
the third group population is not confirmed in high 
significance level.
Therefore, there is a
 chance the third log-normal is not needed. 
This third population is the shortest duration GRB group.
However we already know from the BATSE sample the short 
duration group is exist.  Therefore from
the SAX data,  using only the duration information
we can confirm the existence
of the short bursts with only  95-96\% significance level.

\section{Discussion}

There is no question that the short bursts 
exist, since many satellites identified short
(less than two seconds long) and spectrally harder bursts.
In this paper we are using only the duration
information of the bursts.
The ML method was not able to identify
all the three subgroups in the BeppoSax data at a high
significance level (once again, using only $T_{90}$).
The third group in our analysis was the shortest in
duration. As \cite{fr08} 
pointed out short bursts are less pronounced and displaced toward
higher durations with respect to BATSE. This discrepancy is due to the lower efficiency of
the GRBM trigger system to short GRBs, which, for almost the entire BeppoSax
mission duration, used 1 s as short integration time.
More details can be seen in 
 \cite{band03,g01,fr97}.

However, in the duration distribution two components
were found at a very high significance level.
Surprisingly, analyzing the duration distribution
observed by BeppoSax GRBM one can find
the long and the intermediate duration population.
The short population can be seen only with low
significance level.

\section{Conclusions}

In the BATSE data many scientific groups identified
three types of bursts; short hard, long soft and
intermediate duration very soft bursts
\citep{muk98,ho98,hak00,bala01,rm02,ho02,hak03,ho06,ch07}.
In the Swift \cite{ho08} and RHESSI  \cite{rip07,rip09}
data recent works identified the
same group structures.

In this paper we analyzed a fourth data set - observed
by the BeppoSAX satellite. The duration distribution
cannot be well fitted with one Gaussian component.
The two Gaussian fit is much better and
the three Gaussian fit is better only at a
 95-96\%  significance level.
However, we surely know from previous works
that short (third group in our analysis) bursts exist.

The existence of the intermediate type burst
is still not widely accepted.
In this paper we showed that in order to fit the BeppoSAX data
duration distribution a second
component of intermediate duration is needed.
 However, the physical
existence of the intermediate GRBs can be still 
questionable. \cite{hak03} argued that
the statistical existence of the intermediate
group in the BATSE sample is caused by instrumental effects.
Instrumental biases can  play a relevant role in shaping
the observed duration distribution and the different fractions of short GRBs
observed in the various catalogues reflects this. In the case of the GRBM,
an important bias is given by the relatively long (1 s) short integration
time adopted throughout most of the mission lifetime  \cite{fr08}.
Due to this, the claim for a physically separate class of intermediate GRBs
requires a word of caution and the upcoming catalogues of new missions, such
as Fermi/GBM, will help to clarify this issue.

\acknowledgments
This research was supported in part through  OTKA T048870 and   
K 77795 grant and Bolyai Scholarship.
 Some remarks of J. \v{R}\'{\i}pa are appreciated. 
The author also thanks the referee for useful comments 
and suggestions that improved the paper. 
Especially, the author acknowledge 
useful communications with David Band, who was a
great scientist and a good friend,
and dedicate this paper to his memory.

\nocite{*}
\bibliographystyle{spr-mp-nameyear-cnd}
\bibliography{biblio-u1}

\begin{thebibliography}{}


\bibitem[Balastegui et al. 2001]{bala01}
Balastegui, A., Ruiz-Lapuente, P. \& Canal, R. 2001, MNRAS, 328, 283

\bibitem[Band 2003]{band03} Band, D. L. 2003, Apj, 588, 945

\bibitem[Chattopadhyay et al. 2007]{ch07} Chattopadhyay, T., et al. 2007
ApJ, 667, 1017

\bibitem[D'Agostini 2003]{da03} D'Agostini, G. 2003, 
Bayesian Reasoning in Data Analysis: A Critical Introduction 
(World Scientific, Singapore)

\bibitem[D'Agostini 2004]{da04} D'Agostini, G. 2004, 
eprint arXiv:physics/040308

\bibitem[Frontera et al. 1997]{fr97} Frontera, F., et al. 1997, 
A\&AS, 122, 357

\bibitem[Frontera et al. 2009a]{fr08} Frontera, F., et al. 2009, 
ApJS, 180, 192

\bibitem[Frontera et al. 2009b]{fr09} Frontera, F., et al. 2009, 
http://adsabs.harvard.edu/ abs/2009yCat..21800192F

\bibitem[Guidorzi et al.  2001]{g01} Guidorzi, C.,  et al. 2001,
in Gamma Ray Bursts in the Afterglow Era, ed. E. Costa, F. Frontera, \& J. Hjorth, 43

\bibitem[Hakkila et al. 2000]{hak00} Hakkila, J., et al. 2000, ApJ, 538, 165

\bibitem[Hakkila et al. 2003]{hak03} Hakkila, J., et al. 2003,
ApJ, 582,320


\bibitem[Horv\'ath 1998]{ho98} Horv\'ath, I. 1998, ApJ, 508, 757


\bibitem[Hor\-v\'ath 2002]{ho02} Horv\'ath, I. 2002, A\&A, 392, 791

\bibitem[Horv\'ath et al. 2006]{ho06} Horv\'ath, I., et al. 2006, A\&A, 447, 23

\bibitem[Horv\'ath et al. 2008]{ho08} Horv\'ath, I., et al. 2008, A\&A, 489, L1-L4

\bibitem[Kendall \& Stuart 1976]{KS76}
Kendall, M. \& Stuart, A. 1976, The Advanced Theory
of Statistics (Griffin, London)

\bibitem[Kouveliotou et al. 1993]{kou93} Kouveliotou, C., et al. 1993, 
ApJ, 413, L101

\bibitem[Mukherjee et al. 1998]{muk98} Mukherjee, S., Feigelson, E.D., Babu, 
G.J., Murtagh, F., Fraley, C. \& Raftery, A. 1998, ApJ, 508, 314

\bibitem[Rajaniemi \& M\"ah\"onen 2002]{rm02} Rajaniemi, H.J.,
\& M\"ah\"onen, P. 2002, ApJ, 566, 202

\bibitem[\v{R}\'{\i}pa et al. 2007]{rip07}    \v{R}\'{\i}pa J., Hudec R., M\'esz\'aros A., Hajdas W., Wigger C., 2007, Il Nuovo Cimento B,  121, 1493 

\bibitem[\v{R}\'{\i}pa et al. 2009]{rip09}    \v{R}\'{\i}pa J., M\'esz\'aros A., Wigger C., Huja D., 
Hudec R.,  Hajdas W.,  2009,  A\&A, 498, 399


\end{thebibliography}

\end{document}